\documentclass[twocolumn,aps,prb,graphicx,showpacs]{revtex4}
\usepackage{amsfonts}
\usepackage{amsmath}
\usepackage{amssymb}
\usepackage{graphicx}
\usepackage{epsfig}
\usepackage{epstopdf}
\begin{document}

\title{
Spin dynamics and violation of the fluctuation dissipation theorem
in a non-equilibrium ohmic spin boson model. }
\author{Aditi Mitra and A. J. Millis\\Department of Physics, Columbia University\\
538 W. 120th St, New York, NY 10027}
\date{\today}

%\maketitle

\begin{abstract}
We present results for the dynamics of an impurity spin coupled to a
magnetic field and to two ohmic baths which are out-of equilibrium
due to the application of a bias voltage. Both the non-equilibrium
steady state and the rate constants describing the approach to
steady state are found to depend sensitively on the relative
strengths of a magnetic field and a voltage dependent decoherence
rate. Computation of physical quantities including the frequency
dependent ratio of response to correlation functions and the
probabilities of the two spin states allows the extraction of
voltage dependent effective temperatures. The temperatures extracted
from different quantities differ from one another in magnitude and
their dependence on parameters, and in general are non-monotonic.
\end{abstract}


\maketitle

A fundamental question in quantum condensed matter physics is
understanding properties of non-equilibrium many body systems, some
examples being the Kondo effect in quantum dots \cite{neqex1},
ultra-cold gases with rapidly tunable interactions \cite{neqex2},
and strongly driven optical lattices \cite{neqexp3}. While there are
a variety of non-perturbative techniques in place to study
equilibrium systems, these methods cannot be extended to
non-equilibrium systems in a straightforward way \cite{neqrev}. The
experimental accessability \cite{neqex1,neqex2,neqex3} of the
non-equilibrium regime of strongly correlated quantum many body
systems gives rise to the need for developing the formalism further.

Many body systems driven out of equilibrium are known
\cite{neqKth,mamprl} to acquire a steady state that may be quite
different in character from their ground state properties, with the
details of the steady state depending on the nature of the
correlations, as well as on the way in which the system is driven
out of equilibrium. One may characterize a system in steady state by
the response function $\chi(t-t^{\prime})$ describing changes
induced by weak external probes, and by the correlation function
$S(t-t^{\prime})$ describing the probabilities of observing various
configurations of the system. An important and still incompletely
understood issue is the manner in which $\chi$ and $S$
characterizing a non-equilibrium system differ from those describing
an equilibrium one. In particular there has been considerable
interest \cite{fdt1,fdt2} in the possibility of establishing a
generalized fluctuation-dissipation theorem relating $\chi(\omega)$
to $S(\omega)$ and thereby characterizing the departures from
equilibrium in terms of an effective temperature.  Several systems
have been identified \cite{fdt2} where such a generalized
fluctuation dissipation theorem is found to hold, with the extracted
temperature often sensitive to the observables being studied.

In this paper we study the dynamics of the out of equilibrium ohmic
spin-boson model. This model describes a two state system (which we
represent in spin notation) with level splitting $2B$ and tunneling
rate $\Delta$, coupled via a coupling $J_z$ to a spin-less resonant
level (creation operator $d^{\dagger}$), which is itself connected
to two leads ($L$ and $R$) that may be at different chemical
potentials. The Hamiltonian is
\begin{eqnarray}
H &=& S_z B + \Delta S_x  + J_z S_z d^{\dagger} d
+ H_{bath} \label{hdef}\\
H_{bath} &=& \sum_{k,\alpha=L,R} \epsilon_{k} c_{k\alpha}^{\dagger}
c_{k\alpha} + \sum_{k,\alpha = L,R} \left( t_{k\alpha}c_{k
\alpha}^{\dagger} d + h.c.\right)\nonumber \\ \label{hb}
\end{eqnarray}
We assume the leads are infinite reservoirs characterized by the
correlators  $\langle c_{k \alpha}^{\dagger} c_{q \beta}\rangle
=\delta_{kq} \delta_{\alpha \beta}\left( e^{\beta (\epsilon_k -
\mu_{\alpha})} + 1\right)^{-1}$, and a non-equilibrium state occurs
when $\mu_L -\mu_R = V \neq 0$. Crucial parameters of the model are
the left and right channel phase shifts $\delta_{L,R}$ defined by
$\tan{\delta_{L}} = \frac{a_L J_z}{1 - i sgn(V) a_R J_z} $,
$\tan{\delta_{R}} = \frac{a_R J_z}{1 + isgn(V) a_L J_z} $ with
$a_{L,R}=\frac{\Gamma_{L,R}}{\left(\Gamma_L + \Gamma_R\right)^2}$,
with $\Gamma_{L,R} = \pi \rho t_{L,R}^2$ and $\rho =
\frac{dk}{d\epsilon_k}$. We will study properties of $H$ at $T=0$
but out of equilibrium ($V\neq 0$) working to leading nontrivial
order in $\Delta$ but to all order in $J_z$. The conditions under
which perturbation theory is justified will be discussed below. We
will find the time scales characterizing the approach to steady
state, the response and correlation functions and the generalized
fluctuation dissipation ratio.
\begin{figure}
\includegraphics[width=2.0in]{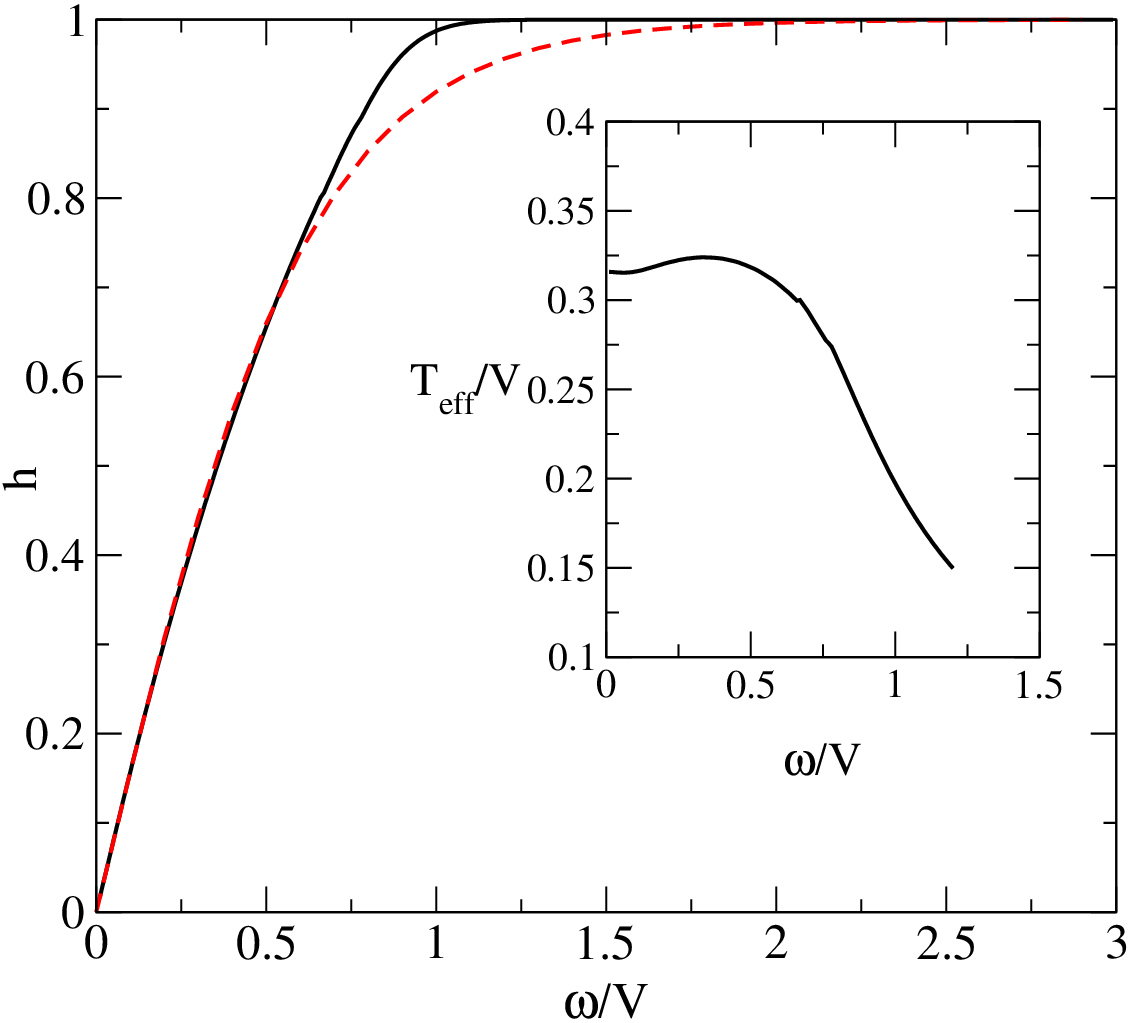}
\caption{Main panel: Effective distribution function derived from
Eq.~\ref{hw} at $B=0.05 V$. The dashed line is
$\tanh\left(\frac{\omega}{2 T_{eff}^h}\right)$. At equilibrium
$h(\omega) = sgn(\omega)$. Inset: Plot of $T_{eff}(\omega,V) =
\frac{\omega}{2 tanh^{-1}h}$ which shows a rapid crossover from
non-equilibrium behaviour ($T_{eff} \sim V$) to equilibrium
behaviour ($T_{eff} = 0$) at $\omega \sim V/2$.} \label{dist}
\end{figure}

In order to study the spin-dynamics, the appropriate starting point
is the density matrix for the full Hamiltonian
\begin{eqnarray}
\frac{d\rho(t)}{dt} = -i \left[H,\rho(t)\right] \label{rhoeom}
\end{eqnarray}
from which the density matrix $\rho_S$ for the local spin is obtained from
taking a trace over the electronic degrees of freedom.
\begin{eqnarray}
\hat{\rho}_{S} = Tr_{el} \rho
\end{eqnarray}
We adopt a spin language, writing
\begin{equation}
\hat{\rho}_S = \frac{1}{2}\left(1 + S_z \hat{\sigma}_z \right)
\label{szdef}
\end{equation}
and study $S_z$. When $\Delta=0$, the Hamiltonian is exactly
solvable both in and out of equilibrium. For non-zero $\Delta$ one
may expand Eq.~\ref{rhoeom} perturbatively in $\Delta$. The key
object in the analysis is the time-evolution operator separating two
spin-flip processes \cite{yuvand}, $K_{\pm}(t) = Tr_{el}\left[
e^{-iH(\Delta=0,S_z=\pm 1/2)t}e^{iH(\Delta=0,S_z=(\mp 1)/2)t}\right]
= e^{\pm i B t} e^{-C_{\pm}(t)}$  where $C_{+}(t)=C(t)
=\left(C_{-}(-t) \right)^*$ computed from the linked cluster theorem
has the following form at zero temperature \cite{mamprl},
\begin{eqnarray}
C(t) = C^{\prime}(t) + i C^{\prime \prime}(t) \nonumber \\
C^{\prime}(t) =
\left( \frac{\delta_{eq}}{\pi}\right)^2 \ln{\left(\xi t\right)} +
\phi^{\prime}(V t)
\label{Qreq} \\
C^{\prime \prime} = \frac{\pi}{2}
\left(\frac{\delta_{eq}}{\pi}\right)^2sgn(t) + \phi^{\prime
\prime}(V t) \label{Qieq}
\end{eqnarray}
Here $\xi$ is a short time cut-off, $\delta_{eq}= \delta_L +
\delta_R = \arctan\left(\frac{J_z}{\Gamma_L + \Gamma_R}\right)$ is
the equilibrium phase shift, and $\phi(Vt)$ is a function describing
the crossover from the short-time ($Vt \ll 1$) equilibrium behaviour
characterized by \cite{xrayeq} $C(t) = \left(
\frac{\delta_{eq}}{\pi}\right)^2 \left[ \ln{\left(\xi t\right)} +
i\frac{\pi}{2} sgn(t)\right] $, to the long time ($Vt \gg 1$)
non-equilibrium behaviour characterized by \cite{Ng} $C(t) = \left(
\frac{\delta_{L}^2 + \delta_R^2}{\pi^2}\right) \left[ \ln{\left(\xi
t\right)} + i\frac{\pi}{2} sgn(t)\right] + \Gamma_{neq} t $ with
$\Gamma_{neq} = V\frac{|\delta_L^{\prime \prime}- \delta_R^{\prime
\prime}|}{2\pi}$. Correct treatment for the intermediate and short
time behaviour of $\phi$ is essential for obtaining correct results
for $\chi(\omega), S(\omega)$. A general analytic expression for
$\phi$ does not exist, here we use perturbation theory to third
order in $J_z$ to obtain \cite{mamprl},
\begin{eqnarray}
\phi(Vt)&=\frac{|\delta _{L}^{^{\prime \prime }}-\delta
_{R}^{^{\prime \prime }} |}{2\pi } V t \left[ \left( \frac{2%
}{\pi }\right) \left( Si(Vt)-\frac{1-\cos (Vt)}{Vt}\right)\right]
 \nonumber
\\
&- \frac{2Re\delta _{L}\delta _{R}}{\pi ^{2}}
\left[\gamma _{e}-Ci(Vt)+\ln (Vt) \right] \\
&-\frac{2iIm\delta _{L}\delta _{R}}{\pi ^{2}} \left[
\frac{2}{\pi}\int_{0}^{1}du\sin (u Vt)\frac{\left[ \left( 1-u\right)
\ln (1-u)+u\ln u\right] }{u^{2}} \right]\nonumber
\end{eqnarray}

Now let us return to the evaluation of various spin observables. In
terms of the symmetric and anti-symmetric time-evolution operators
$K_{s,a}(t) = Re\left[K_{+}(t) \pm K_{-}(t) \right]$, the equation
of motion for the variable $S_z$ parameterized in Eq.~\ref{szdef} to
leading non-trivial order in $\Delta$ is given by
\begin{equation}
\frac{dS_z}{dt} = -\int^{t}_{0} dt^{\prime} \left[ K_s(t,t^{\prime})
S_z(t^{\prime}) + K_a(t,t^{\prime})\right] \label{szeom}
\end{equation}
The Laplace transform of the two scattering rates $K_{s,a}$ defined
by $ \tilde{K}_{s,a}(\lambda)=\int_{0}^{\infty} dt
K_{s/a}(t)e^{-\lambda t} $ have the form,
\begin{eqnarray}
\tilde{K}_s(\lambda) = \Delta^2\int_0^{\infty} dt e^{-\lambda t}
e^{-C^{\prime}(t)}\cos{B t}
\cos{C^{\prime \prime}(t)} \label{ks}\\
\tilde{K}_a(\lambda) = \Delta^2 \int_0^{\infty} dt e^{-\lambda t}
e^{-C^{\prime}(t)}\sin{B t} \sin{C^{\prime \prime}(t)} \label{ka}
\end{eqnarray}
The above equations capture the effect of two sources of decoherence
on spin dynamics, one is a Korringa type decoherence existing even
in equilibrium, while the second arising mathematically from
$C^{\prime}(t)$ is due to a non-zero voltage and is intrinsically
non-equilibrium.

The solution to Eq.~\ref{szeom} can be written as \cite{wprl85}
\begin{equation}
S_{z}(t) = \frac{1}{2\pi i} \int_{c-i\infty}^{c+i\infty}
\frac{d\lambda}{\lambda} e^{\lambda t} \frac{\lambda S_{z}(0) -
\tilde{K}_{a}(\lambda)} {\lambda + \tilde{K}_{s}(\lambda)}
\label{szt}
\end{equation}
Eq.~\ref{szt} allows straightforward analysis of the long-time
behaviour. As $t\rightarrow \infty$, the integral is dominated by
the pole at $\lambda=0$ and the residue gives $S_z^{\infty}=
S_z\left(t \rightarrow \infty\right)=
\frac{-\tilde{K}_a(0)}{\tilde{K}_s(0)}$. Consideration of $S_z(t) -
S_z^{\infty}$ then yields the rate at which the system approaches
steady state. In the small $\Delta$ limit, and if at least one of
$B,\Gamma_{neq}$ is not too small, the result is exponential
relaxation with rate $\Gamma_{rel} = K_s(0)$. The value of
$\Gamma_{rel}$ depends crucially on whether the dominant time scales
in Eq.~\ref{ks} are large or small relative to $V^{-1}$. If both $B$
and $\Gamma_{neq}$ are less than $V$, one finds (for compactness we
write for the symmetric case $t_L = t_R$)
\begin{eqnarray}
&\Gamma_{rel} = \tilde{K}_s(0) \nonumber \\ &=\frac{\pi}{2}
\frac{\Delta^2}{\xi}
\frac{1}{\Gamma\left(\alpha\right)}\frac{\sin{\left[\frac{\pi
\alpha}{2} +
(1-\alpha)\arctan\frac{\Gamma_{neq}}{B}\right]}}{\sin{\frac{\pi
\alpha}{2}}} \left(\frac{\sqrt{B^2 + \Gamma_{neq}^2}}{\xi}
\right)^{\alpha-1} \label{dec}
\end{eqnarray}
where the non-equilibrium exponent $\alpha =
\left(\frac{\delta_L}{\pi}\right)^2 +
\left(\frac{\delta_R}{\pi}\right)^2$. The most interesting situation
is the relatively small phase shift limit, in which $\Gamma_{neq}
\ll V$ and Eq.~\ref{dec} holds even when $B\geq \Gamma_{neq}$,
provided $B\ll V$. For $B\gg V$, one should set $\Gamma_{neq}=0$ in
Eq.~\ref{dec} and replace the non-equilibrium exponent $\alpha$ by
the equilibrium exponent
$\alpha_{eq}=\left(\frac{\delta_{eq}}{\pi}\right)^2$, yielding the
familiar $T=0$ Korringa relaxation.The above results of an
exponential relaxation to steady state is obtained from neglecting
the $\lambda$ dependence of $K_s(\lambda)$, which is justified when
\begin{equation}
\frac{\Gamma_{rel}}{\sqrt{\Gamma_{neq}^2 + B^2}} \sim
\frac{\Delta^2}{\xi^2} \left(\frac{\sqrt{B^2 + \Gamma_{neq}^2}}{\xi}
\right)^{\alpha-2}  \ll 1 \label{cond1}
\end{equation}
and therefore holds in the perturbative in $\Delta$ regime provided
the voltage or the magnetic field is not too small. Analysis similar
to the equilibrium case shows that Eq.~\ref{cond1} is also the
condition for validity of perturbation theory in $\Delta$.

At steady state and in the small $\Delta$ limit, the density matrix
for the full system is an incoherent superposition of spin up times
the electronic state appropriate to spin-up and spin-down times the
electronic state appropriate to spin-down, which may be expressed
follows,
\begin{equation}
\rho = \rho_{S} \otimes \rho_{el} = \begin{pmatrix}\rho_{\uparrow}
\rho^{\uparrow \uparrow}_{el}&0
\\ 0 & \rho_{\downarrow} \rho^{\downarrow \downarrow}_{el}
\end{pmatrix}
\label{rhodef}
\end{equation}
where $\rho_{el}^{\uparrow \uparrow}$ is the density matrix
corresponding to Hamiltonian $H$ with $S_z = 1/2$ and $\Delta=0$.
Likewise $\rho_{el}^{\downarrow \downarrow}$ is the density matrix
for $S_z = -1/2$ and $\Delta=0$, while $\rho_{\uparrow,\downarrow} =
\frac{1}{2}\left(1 \pm S_z\right)$. We now calculate the response
and correlation functions appropriate to this state and also study
the fluctuation-dissipation relation between them.

The correlation function we study is,
\begin{eqnarray}
S_{xx}(t_1,t_2) &=& i \langle \{ S_{x}(t_1),S_{x}(t_2) \}_+\rangle \nonumber \\
&=& Tr\left[ \rho \{ S_{x}(t_1),S_{x}(t_2) \}_+  \right]
\label{corr}
\end{eqnarray}
and the corresponding spin response function is,
\begin{eqnarray}
\chi_{xx}(t_1,t_2) &=& -i\theta(t_1 - t_2) \langle\left[ S_{x}(t_1),S_{x}(t_2)\right] \rangle
\nonumber \\
&=&-i\theta(t_1 - t_2) Tr\left[ \rho \left[ S_{x}(t_1),S_{x}(t_2)\right] \right]
\label{susc}
\end{eqnarray}
where the density matrix $\rho$ is evaluated to leading order in the
spin flip term $\Delta$ and hence given by Eq.~\ref{rhodef}. The
Fourier transform of the imaginary part of the response and
correlation functions are,
\begin{eqnarray}
\chi_{xx}^{\prime \prime}(\omega) &=& \rho_{\uparrow}
\left[I(B+\omega) - I(B-\omega) \right]\nonumber \\
&-&  \rho_{\downarrow} \left[I(-B-\omega)-I(-B+\omega)
\right]\label{imchiwf}
\\
-iS_{xx}(\omega) &=&\rho_{\uparrow}
\left[I(B+\omega) + I(B-\omega) \right]\nonumber \\
&+&  \rho_{\downarrow} \left[I(-B-\omega)+I(-B+\omega) \right]
\label{corwf}
\end{eqnarray}
where
\begin{equation}
I(B) = Re\left[\int_0^{\infty} dt e^{i B t} e^{-C(t)} \right]
\label{Ix}
\end{equation}
and $\frac{\rho_{\uparrow}}{\rho_{\downarrow}} =
\frac{I(-B)}{I(B)}$. We also consider the fluctuation-dissipation
ratio (also mentioned in \cite{hooleyspin})
\begin{equation}
h(\omega) = \frac{\chi^{\prime \prime}(\omega)}{iS_{xx}(\omega)}
\label{hw}
\end{equation}
In equilibrium and at  $T>0$ (when $\phi(0)=1$ and $C(t) =
\left(\frac{\delta_{eq}}{\pi} \right) \ln{\left(\frac{\xi}{\pi
T}\sinh{\pi t T} \right)} )$, $h(\omega) = \tanh\frac{\omega}{2T}$.

\begin{figure}
\includegraphics[width=2.0in]{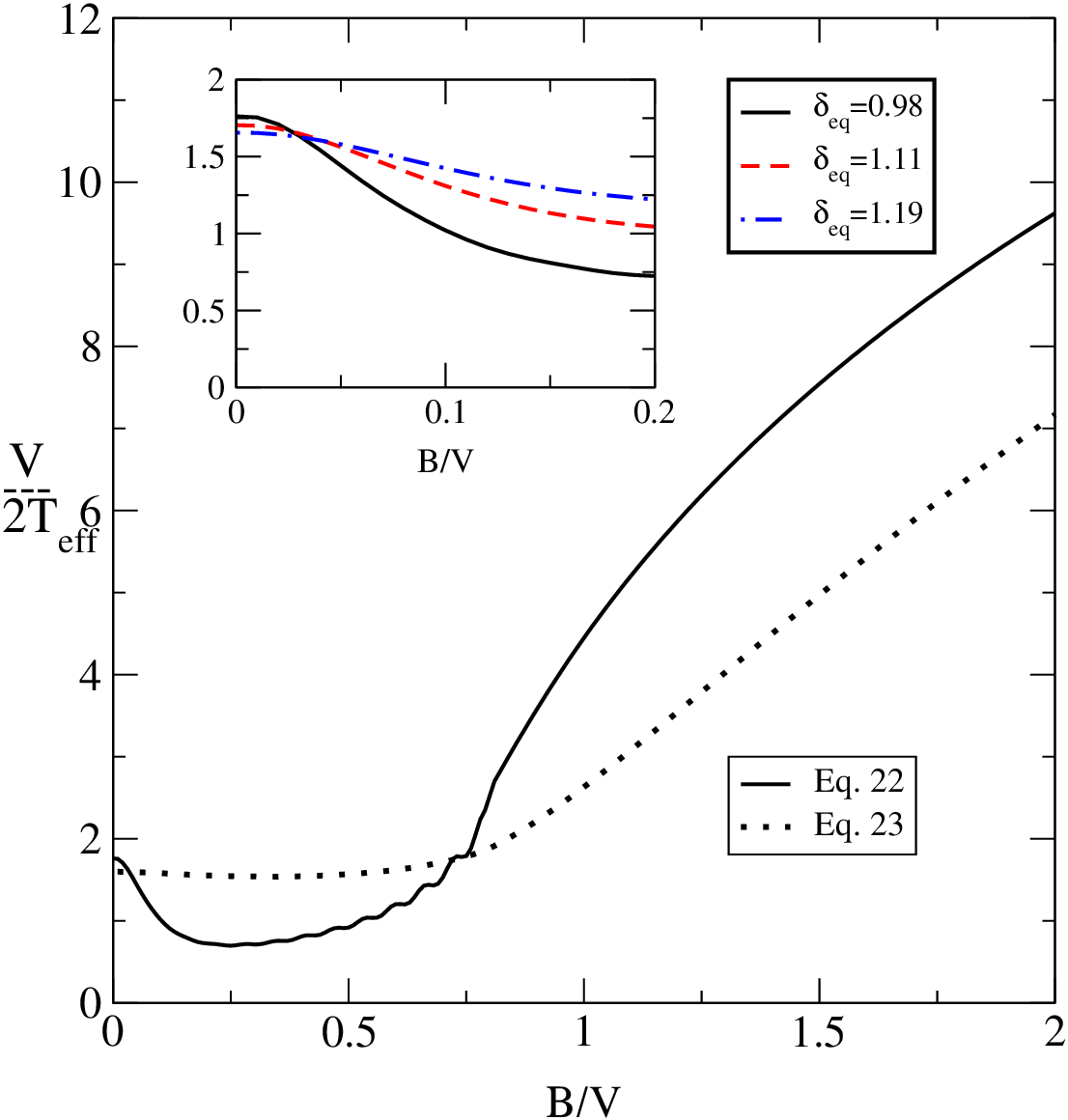}
\caption{Main panel: Low frequency effective temperature defined
from $h(\omega \rightarrow 0)$ (solid line) and population ratios
(dashed line). Note: Non-montonic behaviour with magnetic field.
Inset: Effective temperature (from Eq.~\ref{teff}) as a function of
coupling constant. Note: Increase with $\delta_{eq}$ for $B
\rightarrow 0$, but strong decrease for large $B$.} \label{temp}
\end{figure}

Out of equilibrium (and at $T=0$), $h(\omega)$ has the form shown in
Fig.~\ref{dist} which differs from the equilibrium solution
$sgn(\omega)$. The calculated $h(\omega)$ is not a $\tanh$ function
(compare dashed line), and therefore a generalized fluctuation
dissipation theorem encompassing all frequencies does not exist.
However we may define a frequency dependent effective temperature
via $T_{eff}(\omega) = \frac{\omega}{2 tanh^{-1}h}$. This function
is plotted in the inset of Fig.~\ref{dist} and is seen to have a
strong $\omega$ dependence and is indeed not monotonic. For $\omega
< V/2$, $T_{eff}$ is seen to be of the order of $V$ and to depend
weakly on $\omega$. For $\omega > V/2$, $T_{eff}$ drops sharply and
at high $\omega$ approaches the equilibrium value (here,$T=0$).The
results presented in Fig.~\ref{dist} show that no unique definition
of "non-equilibrium effective temperature" exists, the value
obtained depends on the quantity examined. Two obvious definitions
are (i) from the $\omega \rightarrow 0$ limit of $h(\omega)$
\cite{glass}, (ii) from the population ratio
$\rho_{\uparrow}/\rho_{\downarrow}$. The effective temperature from
definition (i) is obtained by expanding Eq.~\ref{imchiwf} and
Eq.~\ref{corwf} for small $\omega$,
\begin{equation}
\frac{1}{T_{eff}^h} = \frac{\partial h(\omega)}{\partial
\omega}|_{\omega=0} = \sum_{\sigma=\pm}\frac{\partial
\ln{I(x)}}{\partial x}|_{x=\sigma B}\label{teff}
\end{equation}
while definition (ii) for the effective temperature leads to the
expression
\begin{equation}
\frac{1}{T_{eff}^{\rho}} = \frac{1}{B}
\ln{\frac{\rho_{\downarrow}}{\rho_{\uparrow}}}
\end{equation}
Fig.~\ref{temp} shows the dependence of these two measures of
effective temperature on magnetic field. We see that the two curves
differ in magnitude and in dependence on parameters; the variation
in general is non-monotonic. The inset shows that the magnitude and
field variation of the effective temperature (plots are for
$T_{eff}^{h}$) also depends on coupling constant.

The non-monotonic behaviour as a function of $B/V$ may be understood
as follows. For $B\ll \Gamma_{neq}$ and for $\frac{\delta_L^{\prime
\prime}-\delta_R^{\prime \prime}}{2\pi}\ll 1$ so that
$\Gamma_{neq}\ll V$, the integrand in Eq.~\ref{Ix} is dominated by
$t\sim 1/\Gamma_{neq} \gg 1/V$. In this regime, Eq.~\ref{dec}
applies; from this one sees that the decoherence rate for the spin
increases with magnetic field causing the initial downturn in
Fig.~\ref{temp}. This behaviour may also be understood as
originating from the opening up of an additional scattering channel
on application of a magnetic field that corresponds to the
relaxation of the higher energy spin state by creating particle-hole
excitations in the leads. We make this more precise by studying
Eq.~\ref{Ix} perturbatively in tunneling amplitude ($t_{L,R}$) to
find
%$I(B > 0) = \left(a_L^2 + a_R^2 \right)B + a_L a_R \{ (B+V)
%+ \theta(B-V)(B-V)\} ; I(B<0) = a_L a_R \left(
%V-|B|)\right)\theta(V-|B|)$
\begin{equation}
\frac{1}{T_{eff}^h} = \frac{a_L^2 + a_R^2 + a_L a_R}{\left(a_L^2 +
a_R^2 \right)|B| + a_L a_R \left(|B|+V \right)} - \frac{1}{|B|-V}
\label{teffpert}
\end{equation}
For the special case of symmetric couplings $a_L = a_R$ (which
corresponds to the case in Fig.~\ref{temp}), and for $B\ll V$, one
finds $\frac{1}{2T_{eff}^h} \sim \frac{2}{V}\left(1-\frac{2 B}{V}
\right)$ which captures the initial downturn in the plot for
$T_{eff}$.
\begin{figure}
\includegraphics[width=2.0in]{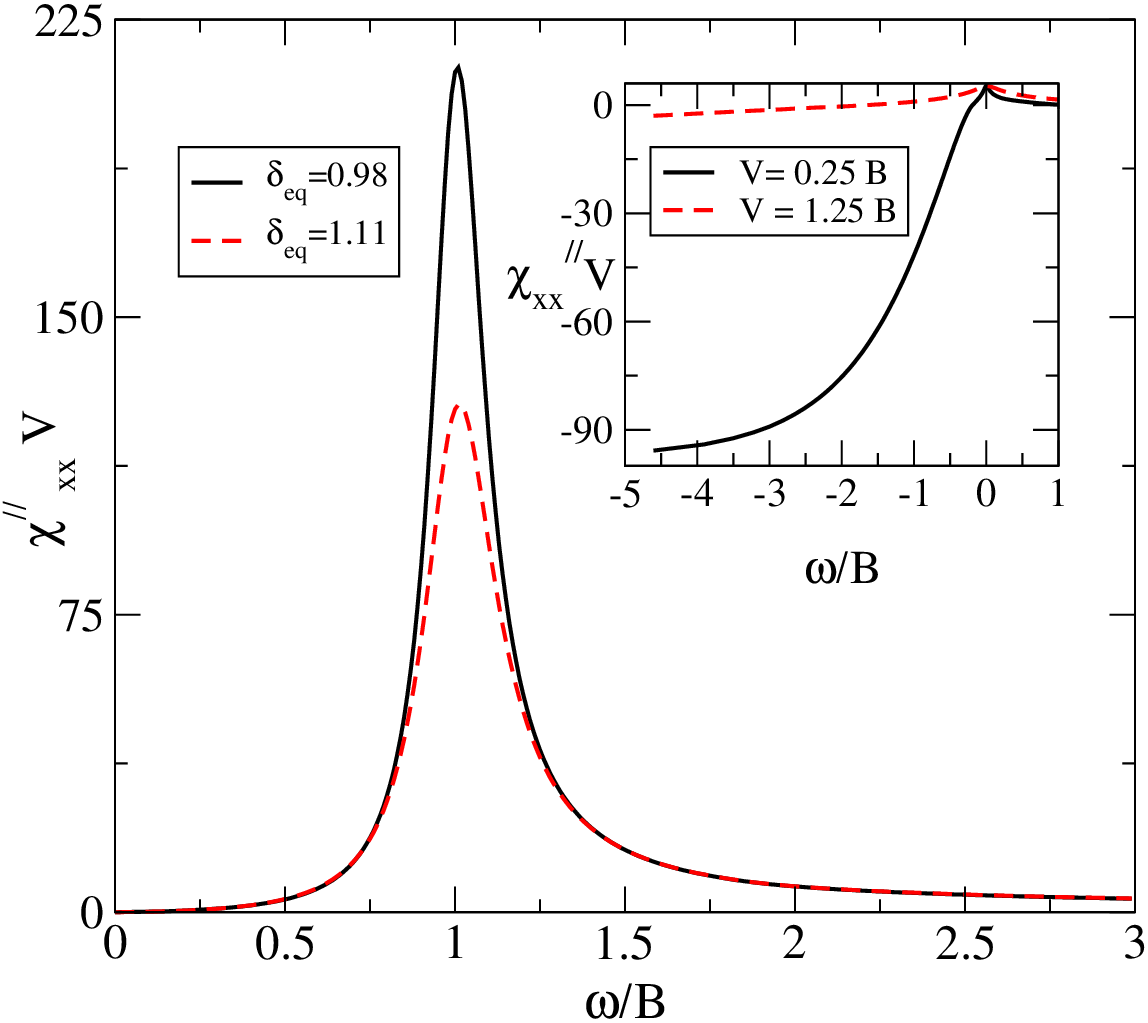}
\caption{Main panel:$\chi^{\prime \prime}(\omega)$ for two different
values of the spin-bath coupling strength $J_z$ and for $B=V$.
Inset: $\chi^{\prime \prime}(\omega)$ on a log-log plot for
spin-bath coupling strength corresponding to  $\delta_{eq} = 0.98$
and for two different degrees of departure from equilibrium.  }
\label{suscV}
\end{figure}

For $B\gg \Gamma_{neq}$ on the other hand the integrals in
Eq.~\ref{Ix} is dominated by $t\sim 1/B \ll 1/V$. In this regime
$T_{eff}$ approaches the equilibrium value of $T_{eff} \rightarrow
0$ and therefore results in an upturn in Fig.~\ref{temp}. The $B\gg
V$ behaviour of the integral $I(B)$ was found \cite{mamprl} to be
$I(B\gg V) \sim \left( \frac{V}{B}\right)^{\frac{B}{V}}$, the
physical significance of which may be understood as follows. $I(B)$
represents the population of the high energy spin state where the
energy for populating it is supplied by the voltage source. The
ratio $n=\frac{B}{V}$ represents the optimal number of bath
electrons that can be transmitted across the voltage source in order
to excite the higher energy spin state, while
$I(B)=\left(\frac{V}{B}\right)^n$ is simply the probability for
doing so. Plugging this expression for $I(B)$ into Eq.~\ref{teff},
the effective temperature is found to approach zero as $T_{eff}^h
\rightarrow \frac{V}{\ln{\frac{B}{V}}}$ in the regime $B/V \gg 1$.

Let us now turn to the discussion of the spin response function
itself. The imaginary part of the response function is plotted for
different coupling strengths and ratio of $B/V$ in Fig.~\ref{suscV}.
The line-shape (main panel: Fig.~\ref{suscV}) in addition to having
the familiar asymmetric form of an x-ray response function
\cite{Ng}, now has a finite weight at $|\omega| < |B|$, which is
forbidden at zero temperatures in equilibrium. The coupling constant
(main panel) and voltage (inset panel) dependence of the broadening
is illustrated in Fig~\ref{suscV}. $\chi^{\prime \prime}(\omega)$ is
linear in $\omega$ for small $\omega$, with a slope that is
inversely related to the long-time relaxation rate of the density
matrix $\Gamma_{rel} = T_{eff}^h$, while at large frequencies
$\omega \gg B$, $\chi^{\prime \prime}(\omega) \sim \frac{1}{\omega
^{1-(\frac{\delta_{eq}}{\pi})^2}}$. These two different frequency
regimes appear as a change in slope of the plots in the inset of
Fig.~\ref{suscV}.

In conclusion, we have studied the non-equilibrium ohmic spin-boson
model including a non-vanishing level splitting and orthogonality
effects exactly. Previous work \cite{hooleyspin} studied the zero
level splitting limit, treating the orthogonality effects
perturbatively. Our results agree with previous results in the
appropriate limit, but provide significant new information including
the non-monotonic effective temperature and the line-shape at
non-vanishing level splitting. The calculated spin dynamics reveal
that the non-equilibrium regime can be quite complex because of the
interplay between various voltage and magnetic field dependent
relaxation mechanisms. While departures from equilibrium are
qualitatively similar to a non-zero temperature (e.g. permitting
sub-threshold absorption of Fig.~\ref{suscV}), the analogy cannot be
pushed too far. The "fluctuation-dissipation" ratio is not a
hyperbolic tangent and indeed is not characterized by a unique
effective temperature (c.f. Fig.~\ref{dist}), and the low-frequency
effective temperature is itself a non-trivial function of the
control parameters (c.f. Fig.~\ref{temp}), and is different
depending on the quantity used to evaluate it. In equilibrium, the
spin boson and Kondo models are related by the simple mapping
$\frac{\delta_{eq}}{\pi} \rightarrow \sqrt{2}\left( 1-
\frac{\delta_{eq}}{\pi}\right) $. Our finding that non-equilibrium
effects enter into different parameters in different ways suggests
that the mapping will not be so simple in the non-equilibrium case.
A direction for future research is to extend the analysis in this
paper to arbitrary number of spin flip processes, and to perform an
Anderson-Yuval-Hamann type renormalization group treatment for the
out of equilibrium spin boson and Kondo models \cite{mmp}. This work
was supported by NSF DMR-0431350.

\end{document}